\documentclass[iop]{emulateapj}
\usepackage{graphicx}

\begin{document}

\shorttitle{A Tidal Stream in the Antlia Dwarf Galaxy}
\shortauthors{Penny et al.}

\title{Tidal interactions at the edge of the Local Group: New evidence for tidal features in the Antlia Dwarf Galaxy}

\author{Samantha J. Penny and Kevin A. Pimbblet}
\affil{School of Physics, Monash University, Clayton, Victoria 3800, Australia}

\author{Christopher J. Conselice}
\affil{School of Physics \& Astronomy, University of Nottingham, Nottingham, NG7 2RD, United Kingdom}

\author{Michael J. I. Brown}
\affil{School of Physics, Monash University, Clayton, Victoria 3800, Australia}

\author{Ruth Gr\"{u}tzbauch}
\affil{Center for Astronomy and Astrophysics, Observatorio Astronomico de Lisboa, Tapada da Ajuda, 1349-018 Lisboa, Portugal}

\author{David J. E. Floyd}
\affil{School of Physics, Monash University, Clayton, Victoria 3800, Australia}

\begin{abstract}
Using deep $B$ band imaging down to  $\sim\mu_{B} = 26$~mag~arcsec$^{-2}$, we present evidence for tidal tails in the Antlia Dwarf galaxy, one of the most distant members of the Local Group. This elongation is in the direction of Antlia's nearest neighbor, the Magellanic-type NGC 3109. The tail is offset by $<10^{\circ}$ from a vector linking the centers of the two galaxies, indicative of interactions between the pair. Combined with the warped disc previously identified in NGC 3109, Antlia and NGC 3109 must be at a small separation for tidal features to be present in Antlia. We calculate that Antlia cannot be completely disrupted by NGC 3109 in a single interaction unless its orbit pericenter is $<6$~kpc, however multiple interactions could significantly alter its morphology. Therefore despite being located right at the edge of the Local Group, environmental effects are playing an important role in Antlia's evolution. 
\end{abstract}

\keywords{galaxies: individual (Antlia Dwarf) --- galaxies: dwarf --- galaxies: interactions --- galaxies: evolution --- Local Group}

\section{Introduction}

The formation of dwarf spheroidal galaxies falls into two categories: either they are old, primordial galaxies that ceased star-formation in the early Universe, or they are relatively young galaxies formed via the morphological transformation of low mass, star-forming galaxies. Through processes such as ram-pressure stripping and tidal interactions \citep{mayer06}, low mass disk galaxies lose their star-forming material and morphologically transform into dwarf spheroidals. However, the degree to which these environmental processes shape the evolution of dwarfs is unknown. By studying nearby low-mass galaxies in both isolation and in pairs/groups, we can potentially gain an insight into the processes that govern low mass galaxy evolution. 

The closest environment in which to study galaxy evolution is the Local Group, with over 40 bona fide members \citep{mateo98}.  Of these, the Antlia Dwarf Galaxy ($M_{V} = -10.4$) is one of the most recent additions. Antlia was noted by \citet{corwin85}, but its distance was determined for the first time by \citet{whiting97} using the tip of the red giant branch as $1.15\pm0.10$ Mpc, which placed it well within the boundaries of the Local Group. More recent determinations of Antlia's distance place it further out than this original estimate, with its distance determined to be between $1.24 \pm 0.07$ Mpc  \citep{blitz2000} and $1.51\pm0.07$ Mpc \citep{piersimoni99}. 

At these distances, it has been hypothesized that Antlia is undergoing significant interaction with its nearest (and more massive) neighbor, the Magellanic-type NGC~3109. \citet{aparicio97} notes that given the distance between the two galaxies may be as large as 180 kpc coupled with a relative of velocity 45 km s$^{-1}$, they are unlikely to be interacting (or even bound) at present.  This supports \citet{pimbblet12} who suggest that the morphology of Antlia is undisturbed to a limiting surface brightness within the central arcmin of $\mu_V = 25.3$ mag arcsec$^{-2}$. However, if the distance difference is lower, (e.g. \citealt{aparicio97} suggest that it could be as low as 28 kpc) then NCG~3109 could have had strong interactions with Antlia (cf. \citealt{vandenbergh99}). This is supported by observations of warping in the disk of NGC~3109 \citep{jobin90,lee03,grebel03,barnes01}.  

In this Letter, we follow up the earlier work of  \citet{pimbblet12} with deep, wide-field observations of Antlia to determine if there is any tidal debris present to unambiguously resolve the questions of whether Antlia has had historic interactions with NGC~3109. This paper utilises data taken using the ESO 2.2~m telescope Wide Field Imager (WFI), providing us with a wide enough field of view to identify any extended structure around Antlia. We find the Antlia dwarf has a two-sided elongation, indicative of a tidal interaction, and conclude this tidal feature is likely the result of interactions with NGC 3109. 

The format of this Letter is as follows. In Section~\ref{obs} we describe our observations, with the analysis of our data presented in \S~\ref{analysis}. The results of this analysis are presented in \S~\ref{res}, and we discuss these results in \S~\ref{discus}, before concluding in \S~\ref{con}.

\section{Observations}
\label{obs}

Using the MPG/ESO 2.2m telescope\footnote{Based on observations collected at the European Organisation for Astronomical Research in the Southern Hemisphere, Chile associated with proposal 69.A-0123(A).}, we obtained  imaging of the Antlia Dwarf  during a WFI survey of the outer regions of the NGC 3175 group of galaxies (Penny et al., in prep). We use the $B$ filter (corresponding $R$ band data will be presented in Penny et al., in prep, but were not taken in sufficiently good seeing for this work\footnote{The $R$ band imaging taken on 2nd February 2011 had variable seeing, ranging from $0.9''$ to $1.7''$, fluctuating throughout the night.}). The total exposure time is 8000~s,  allowing us to reach a low surface brightness of $\mu_{B} = 26$ mag arcsec$^{-2}$, calculated from the rms per square arcsecond. The $B$ band imaging was taken on 3rd February 2011 during dark time in seeing conditions of $0.8''$.

The data are reduced in a standard manner using the \textsc{iraf esowfi} package. A deep twilight flat made from the summation of all twilight flats taken during our 7.5 night observing run is constructed, with each night's calibration observations carefully checked for variations. We find the variation in the background sky of the reduced image is $<0.01$~mag over the area of sky we use in this work.

\section{Analysis}
\label{analysis}

We compare two methods of source detection: \textsc{daophot} and \textsc{sextractor} \citep{bertin96}. 

\begin{figure}
\includegraphics[width=0.8\textwidth]{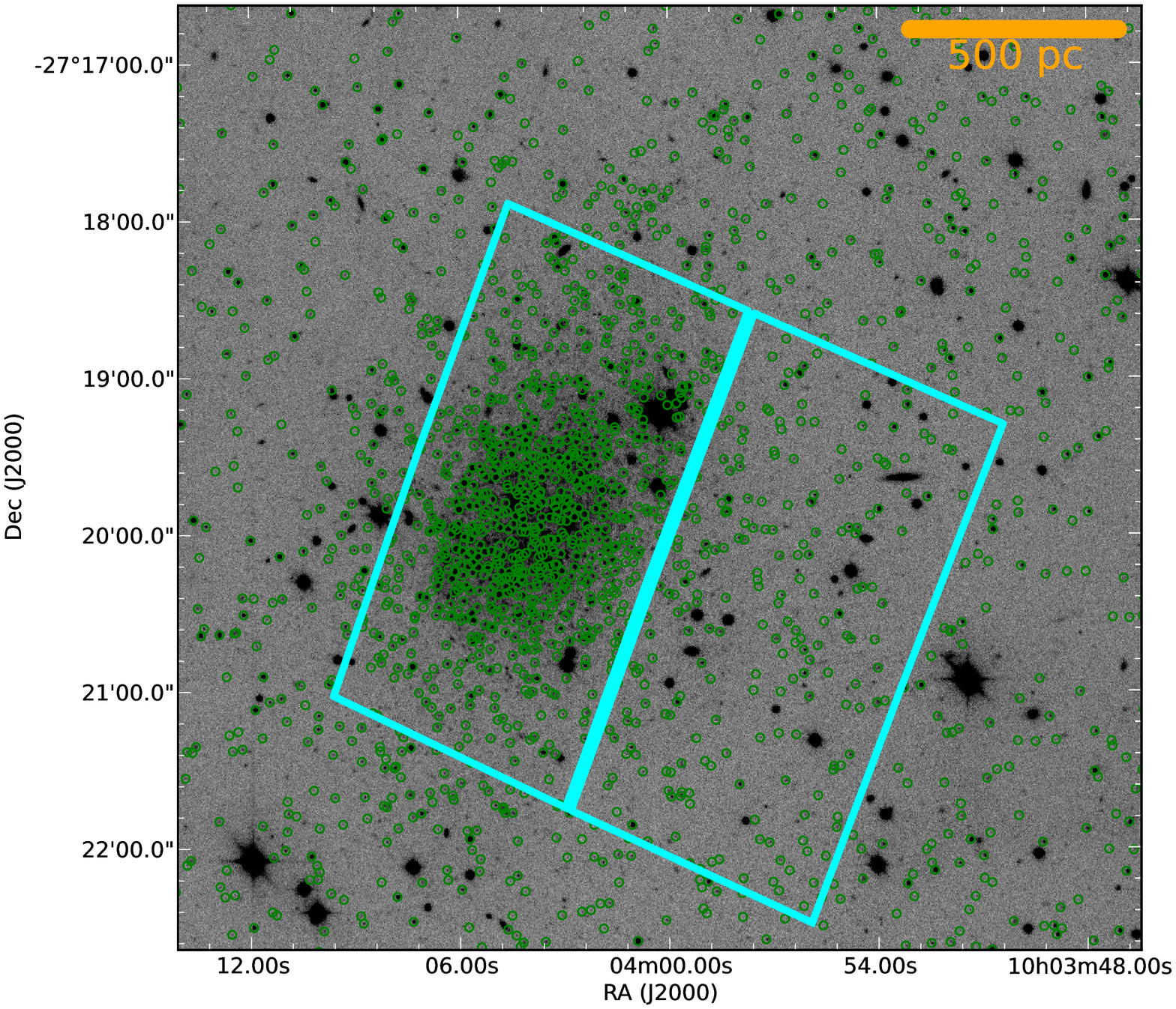}
\caption{WFI image of Antlia. Overplotted are the positions of all objects detected using \textsc{iraf daophot}. The \textit{HST} ACS footprint is shown for comparison.The northern tidal stream of Antlia extends beyond the field-of-view covered by a single pointing of ACS. \label{fig1}}
\end{figure}

Using \textsc{daophot}, we detect all objects $3\sigma$ above the noise. We construct the WFI PSF, and then perform PSF-fitting photometry to all objects. Subtracting the psf-fitted stars from the image of Antlia reveals a low surface brightness feature consisting of unresolved stars. A chi-squared cut of 2 was applied to our catalogue, as was a sharpness cut of $-1 < \textrm{sharp} <1$. We reject sources brighter than the tip of the red giant branch to decrease the likelihood of foreground contamination. The positions of the remaining \textsc{daophot} detections is shown in Fig.~\ref{fig1}. 

Using \textsc{sextractor}, we identify all objects 8 contiguous pixels (pixel scale $0.238''$) in size 0.4$\sigma$ per pixel above the sky background. This results in a minimum S/N of $\sim3$ for sources in our \textsc{sextractor} catalog, with a faint-magnitude cutoff of B=27.5 applied to our catalog. \textsc{sextractor} resulted in 2356 detections vs. 1562  for \textsc{daophot}, due to \textsc{sextractor} detecting unresolved sources missed by \textsc{daophot}. Therefore we prefer to use the results from \textsc{sextractor} in our analysis of local stellar density. 

Any tidal features present in Antlia will be more apparent in the sparser outer regions of the galaxy, where \textsc{sextractor} {\emph is} able to resolve stars as individual sources. This occurs at local densities $<150$~sources~arcmin$^{-2}$. To allow us to detect faint sources in Antlia and a possible tidal stream, we turn the cleaning parameter off, requiring the diffraction spikes of bright stars to be manually cleaned.

When using \textsc{sextractor}, we use a magnitude cut $21.5 <  B <27.5$. This cut rejects all objects $\sim2$ mag brighter than the $B$ band tip of the RGB to account for bright blended stars in the center of the galaxy.  We do not remove background galaxies from our catalog. Apart from a few large galaxies, separating background galaxies from the stellar population of Antlia is difficult without high resolution or multi-band imaging, with archival \textit{Hubble Space Telescope (HST)} Advanced Camera for Surveys (ACS) imaging not covering the full extent of Antlia.

\section{Results}
\label{res}

\subsection{Structure}

\begin{figure}
\includegraphics[width=0.48\textwidth]{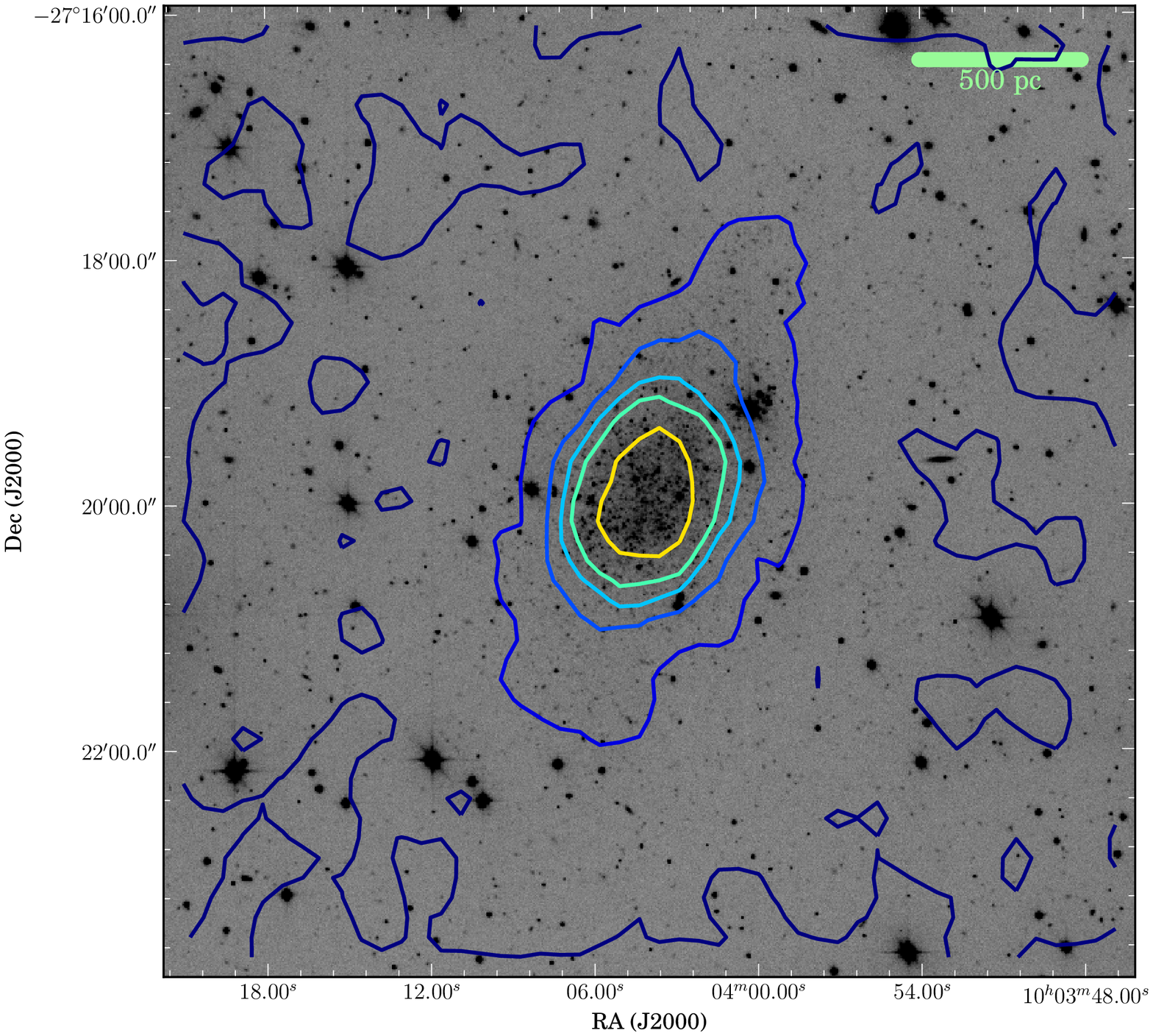} \includegraphics[width=0.45\textwidth]{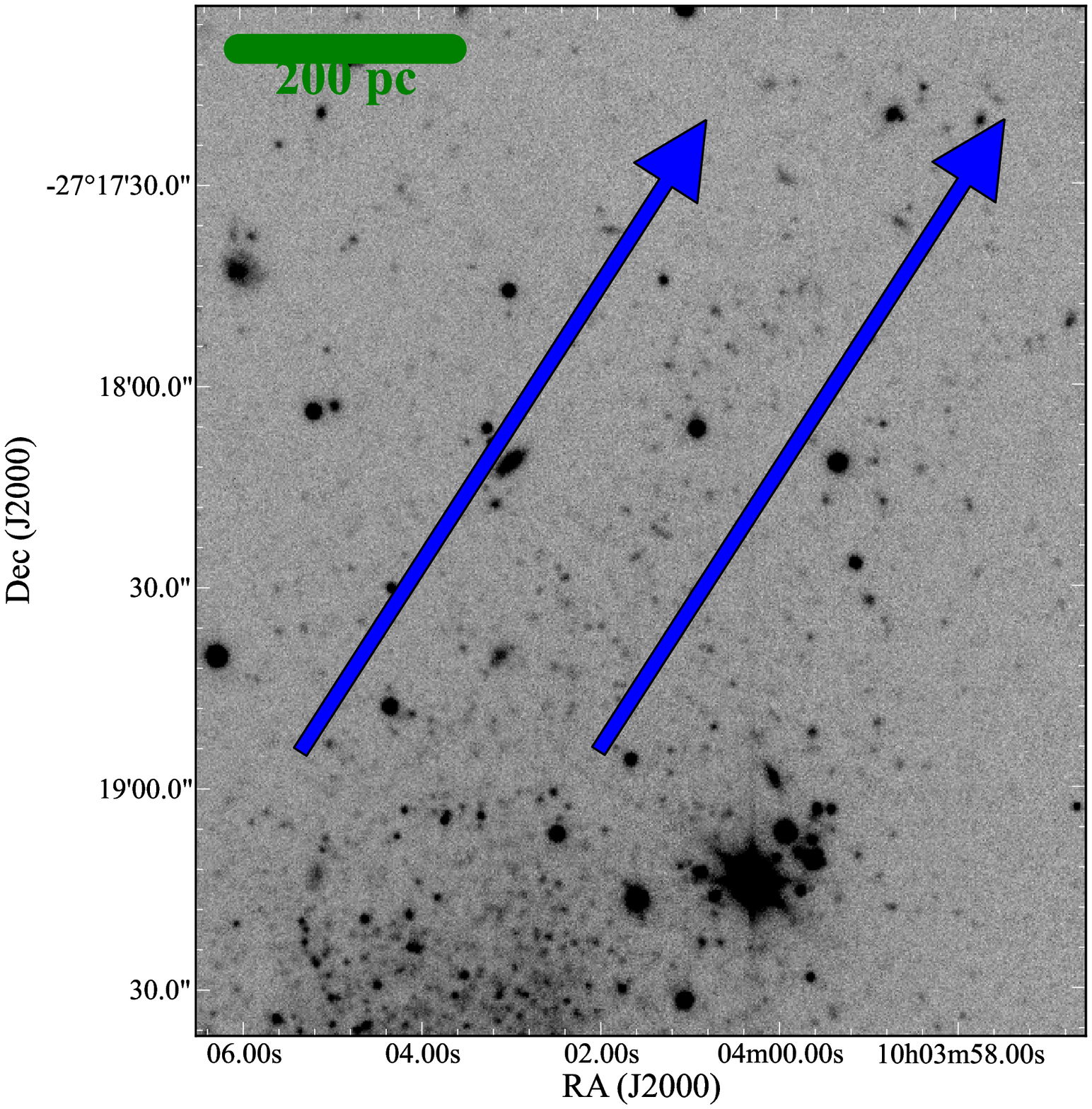}
\label{contours}

\label{tail}
\caption{The left-hand plot shows an ESO 2.2m WFI B band image of the Antlia Dwarf. Contour levels of 15, 50, 100, 150, 200 and 300 sources~arcmin$^{-2}$ are plotted. The contour of 50 sources ~arcmin$^{-2}$ highlights the elongation of the galaxy. The right-hand plot is a $ 2.2' \times 2.6'$ cutout of the northern tidal stream (pointing towards NGC 3109). The blue arrows border the brightest region of the tidal tail, and are offset from a vector linking the centers of Antlia and NGC 3109 by $<10^{\circ}$. \label{fig2}}
\end{figure}

 We examine the structure of the Antlia dwarf to look for signs of tidal interactions. A deep image of the Antlia dwarf is shown in Fig.~\ref{fig2}, with a cutout of the tail also shown. The central $1'$ of the galaxy is easily seen, with a more diffuse distribution of stars surrounding this central concentration. An elongated low surface brightness feature is apparent to the NW of the galaxy's center, suggesting that the structure of Antlia is being tidally influenced (cf.\ Lee et al.\ 2003). 

\begin{figure}
\plotone{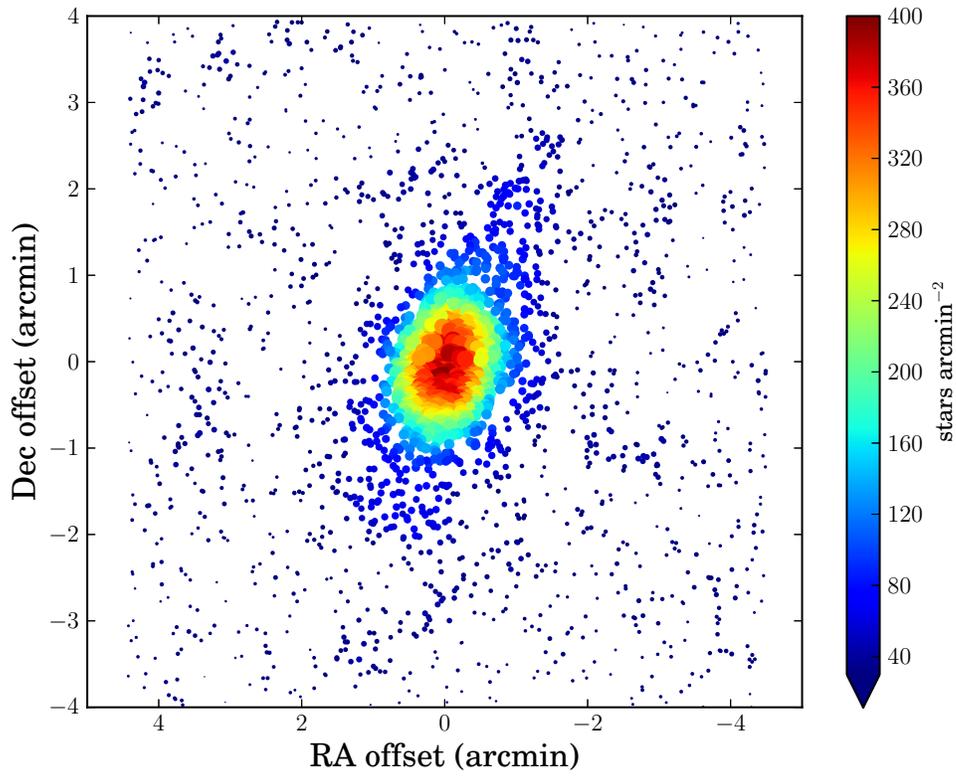}
\caption{The $\alpha$, $\delta$ offset from the center of Antlia for all sources identified in our imaging using \textsc{sextractor}. The points are colored with respect to their local density, with the red points representing regions of highest density, and the darkest blue points representing points of lowest density.\label{fig3}}
\end{figure}

To highlight this low luminosity stream, we examine the local stellar density near each \textsc{sextractor} detection. Low surface brightness features associated with the galaxy would be expected to have a higher local stellar density than the foreground population. The local density near each star is calculated as the number of sources in a $20''$ radius around each detection. $20''$ corresponds to $\sim$120~pc at the distance of Antlia (1.3~Mpc). We display this local density in Fig.~\ref{fig3}. Local densities less than 50 sources~arcmin$^{-2}$ are represented by dark blue points. The elongated feature to the NW is clearly visible in this plot, with elongation also seen on the opposite side of the galaxy. This two-sidedness is consistent with a tidal feature, with the feature more elongated to the NW.   

For dwarf satellites, the inner dense regions of the tidal tail (and therefore most easily detected) will point radially towards the host galaxy (e.g. \citealt{klimentowski09}).  The nearest galaxy to Antlia is the Magellanic-type dwarf irregular NGC 3109, located at $\alpha=$10:03:06.9, $\delta=-$26:09:34, with the Antlia Dwarf located at $\alpha=$10:04:04.1 $\delta=-$27:19:51.6. Therefore we look for elongated structure in this direction. Contours of local stellar density are displayed in Fig.~\ref{contours}. The innermost isodensity contour plotted (286 sources~arcmin$^{-2}$) is fairly circular, with the outer contours exhibiting elongation towards NGC 3109.  The tidal feature is offset by $<10^{\circ}$ from a vector linking the center of Antlia to NGC 3109. The direction of this elongation is highly indicative of tidal interactions between Antlia and NGC 3109.

\subsection{HST colours}

We utilise $HST$ imaging taken from the ANGST database \citep{dalcanton09} to provide essential color information to trace the inner regions of the tidal tail. The ANGST catalogue was manually cleaned of galaxies and any remaining diffraction spikes via visual inspection of the ACS imaging. This cleaned HST catalogue was then matched to our WFI detections to provide color information. Objects that were poorly matched in magnitude in a plot of $HST$ F606W magnitude vs. WFI $B$ magnitude were then discarded, as they likely correspond to miss-matched detections between the two surveys.

\begin{figure}
\plotone{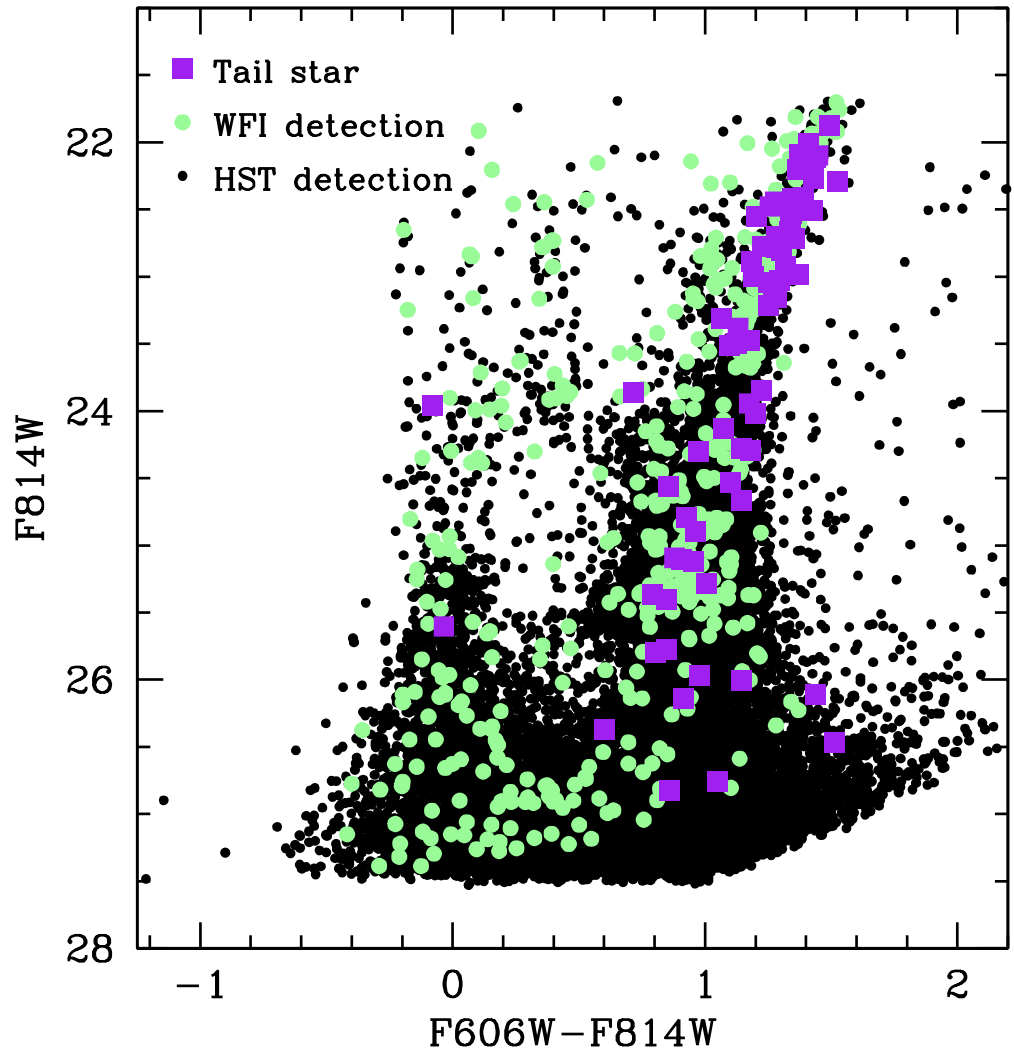}
\caption{Color-magnitude diagram for Antlia. The photometry is taken from the ANGST survey. Black points are all stars in the $HST$ imaging after our cleaning routine, green points are all stars detected in our WFI imaging, and purple squares are objects located in the tidal tails of the galaxy. The majority of objects in the tidal tail lie on the red-giant branch for Antlia, indicative of galaxy membership opposed to foreground contamination. \label{fig4}}
\end{figure}

We then examine the colour-magnitude relation for the cross-matched stars using the colour information provided by HST (Fig.~\ref{fig4}).  All star brighter than the tip of the red giant branch were removed to decrease the likelihood of foreground contamination. We use $F814W = 21.687 \pm 0.049$ from \citep{pimbblet12} as the value of the TRGB. The remaining objects lie on the colour-magnitude relation, with the majority of objects falling on the red giant branch, indicating their membership of the Antlia dwarf galaxy.

\section{Discussion}
\label{discus}

We argue that the elongated features seen in Antlia must be due to recent tidal interactions with NGC 3109. The tails we have identified are offset from a vector linking the centers of NGC 3109 and Antlia by $<10^\circ$, consistent with the predicted orientation of dwarf galaxy tidal tails in the $N$-body simulations of \citet{klimentowski09}. Furthermore, there are no other galaxies in a radius of 100~kpc and within $\pm 500$~km~s$^{-1}$ of Antlia. Despite being located at the edge of the Local Group where we might expect environmental effects to be relatively weak, Antlia is nevertheless undergoing environmentally driven evolution via tidal interactions.

In addition to the tidal extensions around Antlia that we have identified, previous authors have noted features in NGC 3109 that show it is likely interacting with Antlia. NGC 3109 and Antlia  have an angular separation of $1.18^{\circ}$ (a projected distance of 28~kpc), and have near identical distances of $\sim1.3$~Mpc.  Furthermore, the disk of NGC 3109 exhibits an extended low surface brightness structure \citep{hidalgo08}, the result of either mergers or tidal interactions with a neighboring galaxy. This warping of the optical disk is complemented by \citet{barnes01}, who identify H\textsc{i} to the SW of NGC 3109 at 360~km~s$^{-1}$, the radial velocity of Antlia.  Given the numerous pieces of evidence that have now been assembled, Antlia must be the cause of these tidal features in NGC 3109. 

The brightest tidal features in dwarf galaxies are expected to be short lived, and will thus dissipate relatively quickly. A fraction of stars in the tails at small galactocentric radii will fall back towards the galaxy, with stars at larger radii forming streams of tidal debris  tracin the orbit of the dwarf. This result implies that Antlia and NGC 3109 cannot be at a large separation for tidal features to remain detectable in our imaging. Following the simulations of \citet{klimentowski09}, the inner tidal tails of Milky Way dSphs will be at their densest and most easy to detect a few hundred Myr after pericenter, with lifetimes $<1$~Gyr. Although the satellites of the Milky Way have tidal streams that can be long lived ($>1$~Gyr), they are typically of very low surface brightness (e.g. $\mu_{V} >29$~mag~arcsec$^{-2}$ for Ursa Minor, \citealt{md01}), and would therefore be undetectable in our imaging.

While \citet{pimbblet12} suggest that Antlia probably has not undergone recent merger events with other galaxies, we note the tidal tail we identify in this work extends beyond the survey area of a single pointing of $HST$ ACS by $\sim 40''$. Indeed, the $HST$ imaging of Antlia only shows the outer regions of the galaxy to the E, where we see no evidence for elongation. Therefore the elongation we identify in the direction of NGC 3109 would not have been picked up in their asymmetry determination for Antlia.  Examining this tidal tail in detail is the next step in this work, with deep, multi-band imaging taken in stable seeing (else space-based imaging) required for an analysis of its stellar population. Further observations of the extended structure in NGC 3109 could be used to determine if some of this structure originated in Antlia.

\citet{grebel03} identify Antlia as a transition dwarf galaxy, with a morphology intermediate between a dIrr and dSph. The recent interaction with NGC 3109 has therefore likely removed some of Antlia's H~\textsc{i} gas, and further interactions could shut off the supply of gas for star-formation in the dwarf. The interactions between Antlia and NGC 3109 are likely an example of pre-processing via tidal interactions, whereby a star-forming dIrr is transformed into a dSph in the group environment.

An additional piece of evidence for a recent tidal interaction comes from Antlia's star-formation history. While the outer regions of the galaxy  consists of an old stellar population, low level star formation only persists in the core of the galaxy \citep{pimbblet12}. Its core contains evidence for a major burst of star formation 640 Myr in duration that ceased $\sim200$ Myr ago  \citet{mcquinn10}. Galaxy interactions can strip dwarfs of their star forming material in their outer regions, which could explain Antlia's recent star formation history.  

In the Local Group, dwarf ellipticals that exhibit tidal features are relatively common, with the Sagittarius dwarf elliptical being an extreme example of a tidally disrupting dwarf around the Milky Way. Around M31, NGC 205 and M32  \citep{choi02} both exhibit tidal features. However, with the exception of the Magellanic Clouds, which are an extremely unusual galaxy pair \citep{james11}, dwarf irregulars are typically well isolated (e.g. WLM and IC~1613). Despite being so remote and located at the edge of the Local Group, Antlia is nevertheless undergoing tidal interactions, making it unsuitable as a target for examining internally-driven dwarf galaxy evolution.  Deeper, wide field observations of other Local Volume dwarf galaxies to a low surface brightness will establish how common tidal features in dwarf irregulars are. 

\subsection{Alternative origins for the stellar stream}

 \citet{grebel03} identify Antlia as a transition dwarf with properties intermediate between that of a dIrr and a dSph. In the Local Group, transition dwarfs show little evidence for disks, including Leo A, Sagittarius dIrr and GR8 \citep{grebel03}. Therefore we consider it unlikely that at present times Antlia hosts a stellar disk.

Could this feature be a globular cluster in the process of being disrupted by its host galaxy? Local Group dwarf galaxies are known to have globular cluster systems, with the Fornax dwarf spheroidal for example having five \citep{harris79}. The width of a tidal stream tells us its origin, with the tidal tails of globular clusters have widths $<100$~pc. The northern tail we identify in Antlia has an angular size $\sim60''$-$90''$, corresponding to 400-600~pc at the distance of Antlia. This is comparable to the widths of tidal streams seen for satellites of the Milky Way, with the Orphan Stream (associated with the Ursa Major II dwarf) $\sim600$~pc in width \citep{belokurov07}. Combined with direction of the tails towards NGC 3109, tidal interactions with the more massive galaxy are the likely origin of this stream. 

The \textit{V}-band surface brightness profile is well fit by a double exponential profile \citep{aparicio97}. The elongation we see towards  NGC 3109 could therefore be the result of this two-component structure. Furthermore, Antlia's mixed-age stellar population with young stars located in its center, and older stars in the outer regions could be a core-halo structure. Such a structure is fairly common in dwarfs, with the gas responsible for star formation concentrated to the center of the galaxy. The lack of star formation in the outer regions of Antlia does not necessarily indicate tidal stripping of Antlia's gas by NGC 3109. Deep, high resolution, multi-band imaging is necessary to fully test this. 

\subsection{Interactions with NGC 3109}

Interactions between low mass dwarf galaxies and giant galaxies such as the Milky Way can completely disrupt dwarf galaxies, as evidenced by the numerous streams of stars in the Milky Way's halo, and in the extremely elongated shape of some of the MW's satellites (e.g. Sagittarius). However, would a single interaction between Antlia and a Magellanic-type irregular (such as NGC 3109) be able to completely disrupt Antlia? We investigate the likelihood of such a scenario. 

We calculate the increase in internal energy from an impulsive interaction to the internal binding energy of an Antlia-mass dwarf (10$^{8}$~M$_{\odot}$) interacting with a Magellanic-type galaxy (10$^{9}$~M$_{\odot}$) (eqn.~9 of \citet{cgw01}). We are unable to find an internal velocity dispersion measurement for Antlia in the literature, but a typical $M_{B} = -11$ dwarf will have an internal velocity dispersion of $\sim10$~km~s$^{-1}$.  The relative velocity of the two galaxies is  45~km~s$^{-1}$. We furthermore assume Antlia has a radius $\sim1.0$~kpc, comparable to twice its effective radius of 471~pc presented in \citet{sharina08} and \citet{mcconnachie12}. There will be little increase in Antlia's internal energy for impact parameters $>10$~kpc, and Antlia will not be completely disrupted in a single interaction with NGC 3109 unless it comes within 6~kpc of the more massive galaxy. Therefore it is unlikely that Antlia will be destroyed in a single interaction with NGC 3109. Instead, repeated interactions could alter the morphologies of both galaxies. 

\section{Conclusion}
\label{con}

The presence of an elongated distribution of stars to the NW and SE of Antlia is highly indicative of a recent tidal interaction with NGC 3109. Therefore we argue that NGC 3109 and the Antlia dwarf are currently interacting/merging, contrary to the analysis of \citet{pimbblet12} due to the smaller angular coverage of their ACS observations. Such tidal interactions would rule out a large separation of the two galaxies. While the elongation in Antlia could be a two-component structure, the direction of the tidal tail towards NGC 3109 is indicative of a galaxy-galaxy interaction as the source of Antlia's structure. Antlia therefore represents the ideal target in the Local Group in which to examine environmentally driven galaxy evolution. Given dwarf galaxies are so numerous, interactions between dwarf galaxies such as NGC 3109 and Antlia might be commonplace, and could play a significant role in shaping the morphologies of dwarf galaxies in groups and clusters. 

\acknowledgements

S. Penny acknowledges the support of an Australian Research Council Super Science Postdoctoral Fellowship grant FS110200047.  M.B. acknowledges support from the Australian Research Council via Future Fellowship grant FT100100280. DJEF acknowledge support from the Australian Research Council via Discovery Project grant DP110102174. We also thank Warrick Couch for useful discussion.

\clearpage

\end{document}